\documentclass[aps,prl,reprint,superscriptaddress]{revtex4-1}
\usepackage{graphicx}
\usepackage{amsmath}
\usepackage[dvips]{color}

\begin{document}

\title{Nonlinear mechanics with photonic crystal nanomembranes}

\author{Thomas Antoni}
\email{thomas.antoni@spectro.jussieu.fr}
\affiliation{Laboratoire Kastler Brossel, UPMC-ENS-CNRS, Case 74, 4 place Jussieu, F75252 Paris Cedex 05, France}
\affiliation{Laboratoire de Photonique et de Nanostructures LPN-CNRS, UPR-20, Route de Nozay, 91460 Marcoussis, France}
\author{Kevin Makles}
\affiliation{Laboratoire Kastler Brossel, UPMC-ENS-CNRS, Case 74, 4 place Jussieu, F75252 Paris Cedex 05, France}
\author{R\'{e}my Braive}
\affiliation{Laboratoire de Photonique et de Nanostructures LPN-CNRS, UPR-20, Route de Nozay, 91460 Marcoussis, France}
\affiliation{Universit\'{e} Paris Diderot, 75205 Paris, Cedex 13, France}
\author{Tristan Briant}
\author{Pierre-Fran\c cois Cohadon}
\affiliation{Laboratoire Kastler Brossel, UPMC-ENS-CNRS, Case 74, 4 place Jussieu, F75252 Paris Cedex 05, France}
\author{Isabelle Sagnes}
\author{Isabelle Robert-Philip}
\affiliation{Laboratoire de Photonique et de Nanostructures LPN-CNRS, UPR-20, Route de Nozay, 91460 Marcoussis, France}
\author{Antoine Heidmann}
\affiliation{Laboratoire Kastler Brossel, UPMC-ENS-CNRS, Case 74, 4 place Jussieu, F75252 Paris Cedex 05, France}

\begin{abstract}
Optomechanical systems close to their quantum ground state (QGS)
\cite{QGS_Teufel,QGS_Painter} and nonlinear nanoelectromechanical
systems (NEMS) are two hot topics of current physics research.
Demonstrating the QGS allows to shed new light on quantum coherent
effets in meso- and macroscopic systems, whereas NEMS operated in a
nonlinear regime \cite{rhoads:034001} are used either to demonstrate
the underlying physics of sheer nonlinear effects
\cite{Cleland,Roukes,Kotthaus} or as RF amplifiers, with sensitivity
improvement both at the classical \cite{Buks,Pistolesi} or quantum
noise level \cite{Buks2,Yamaguchi}. As high-reflectivity and low
mass are crucial features to improve optomechanical coupling towards
the QGS, we have designed, fabricated and characterized photonic
crystal (PhC) nanomembranes \cite{Antoni:11}, at the crossroad of
both topics. Here we demonstrate a number of nonlinear effects with
these membranes. We first characterize the nonlinear behavior of a
single mechanical mode and we demonstrate its nonlocal character by
monitoring the subsequent actuation-related frequency shift of a
different mode. We then proceed to study the underlying nonlinear
dynamics, both by monitoring the phase-space trajectory of the free
resonator and by characterizing the mechanical response in presence
of a strong pump excitation. We  observe in particular the frequency
evolution during a ring-down oscillation decay, and the emergence of
a phase conjugate mechanical response to a weaker probe actuation.
Our results are crucial to understand the full nonlinear features of
the PhC membranes, and possibly to look for nonlinear signatures of
the quantum dynamics \cite{Lifshitz}.
\end{abstract}

\maketitle

We focus on the mechanical nonlinear behavior of photonic crystal
nanomembranes developed as end mirrors in a Fabry-Perot
optomechanical cavity \cite{Antoni:11}. Details about the mechanical
and optical properties, as well as the fabrication process of these
membranes are described in previous publications \cite{Antoni:11,
PhysRevLett.106.203902,talneau:061105}. They are $10 \, \mathrm{\mu
m} \times 20 \, \mathrm{\mu m} \times 260 \, \mathrm{nm}$ indium
phosphide slabs suspended over the substrate by four decoupling
bridges (0.5-$\mu$m width and 6- to 12-$\mu$m lengths) located at
the nodes of the mechanical mode of interest \cite{Aspelmeyer}, $5.9
\, \mathrm{\mu m}$ away from the center of the membrane.

\begin{figure}
\includegraphics[width=8cm]{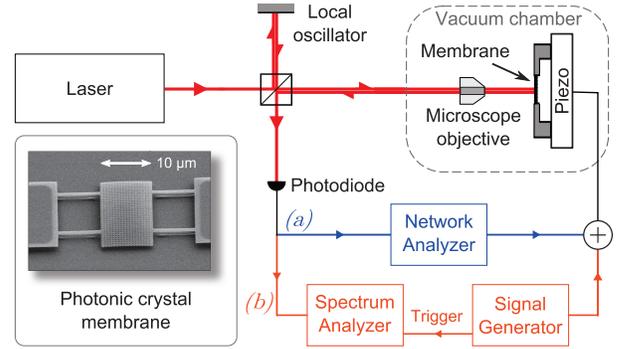}
\caption{\label{Montage} Interferometric setup: we use either a
network analyzer to probe the mechanical response of the membrane
{\it (a)}, or a signal generator and a spectrum analyzer triggered
by the generator to test the ring-down behavior {\it (b)}, or both
of them to perform pump-probe experiments. The inset shows a
typical photonic crystal membrane.}
\end{figure}

To probe the mechanical response of the membrane, the sample is
actuated by a piezoelectric stack, and the membrane displacement
is monitored by a Michelson interferometer (see
Fig.~\ref{Montage}). The sample is operated in a vacuum chamber at
$10^{-2} \, \mathrm{mbar}$ to prevent from air viscous damping and
squeezed film effects. We use three different membranes with
resonance frequencies $\Omega_0/2\pi= 782\ {\rm kHz}$, $1036\ {\rm
kHz}$, and $1057\ {\rm kHz}$, and quality factors $Q\simeq
5\,000$, mainly limited by surface effects.

\begin{figure}
\includegraphics[width=8.6cm]{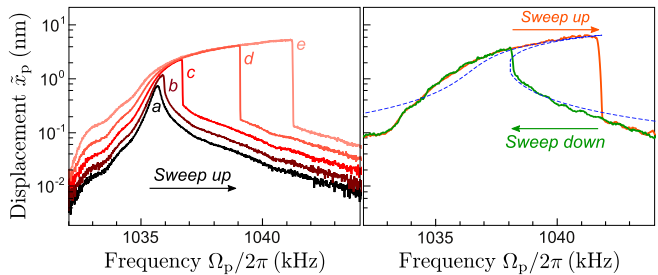}
\caption{\label{membrane} Oscillation amplitude
$\tilde{x}_\mathrm{p}[\Omega_{\rm p}]$ of the membrane as a
function of the piezoelectric actuation frequency $\Omega_{\rm
p}$. Linear to nonlinear transition is clearly visible on left
curves {\it a} to {\it e}, obtained with an upward frequency sweep
of the network analyzer and for increasing actuation powers ($10\
{\rm dBm}$ to $30\ {\rm dBm}$ with a 5-dB step). Curves on the
right are swept either up or down with the frequency generator at
$30\ {\rm dBm}$ and exhibit a typical hysteresis cycle as
expected from the theoretical fit (dashed curve).}
\label{faisceau}
\end{figure}

A network analyzer actuates the sample and monitors the resulting
displacement spectrum.  Figure \ref{faisceau} (left) shows the
forced oscillation amplitude $\tilde{x}_\mathrm{p}[\Omega_{\rm p}]$
as a function of the actuation frequency $\Omega_{\rm p}$, for
different actuation powers. The mode of interest, close to $1\,{\rm
MHz}$ for the membrane used here, exhibits a strong nonlinearity
even for an actuation voltage limited to a few volts applied to the
piezoelectric stack, corresponding to displacements of a few
picometers only for the substrate and at the nanometric level for
the membrane.

The nonlinearity can be accounted for by the Duffing model in which
the resonance frequency has a quadratic dependence with the
displacement $x_{\rm p}(t)$:
\begin{equation}
\ddot{x}_ p(t) + \Gamma
\dot{x}_\mathrm{p}(t) + \Omega_0^2 \left[1 + \beta
x^2_\mathrm{p}(t) \right] x_\mathrm{p}(t) =
\alpha_{p}(t), \label{Eq:Duffing}
\end{equation}
where $\alpha_\mathrm{p}(t) = 2 \tilde{\alpha}_\mathrm{p}
\cos(\Omega_\mathrm{p} t)$ is the driving force,
$\Gamma=\Omega_0/Q$ the mechanical damping, and $\beta$ the
nonlinearity strength.

At first order the forced displacement is mainly monochromatic at
the actuation frequency $\Omega_{\rm p}$, with a Fourier component
given by (assuming $Q\gg 1$)
\begin{equation}
\tilde{x}_\mathrm{p}[\Omega_\mathrm{p}] =
\frac{\tilde{\alpha}_{\rm p}}{\Omega_0^2 \left(1 + \varepsilon
\right) - \Omega^2_\mathrm{p} - i\Gamma\Omega_0}.
\label{Eq:S_x_bista}
\end{equation}
This expression corresponds to the usual lorentzian resonance,
except for a shift of the resonance frequency proportional to the
mean square displacement through the term
\begin{equation}
\varepsilon = 3 \beta\left|\tilde{x}_\mathrm{p}[\Omega_{\rm p}]\right|^2. \label{Eq:epsilon}
\end{equation}
This shift is responsible for the spectra observed in Fig.
\ref{faisceau}, as shown by the theoretical fit on the right
(dashed curve). Although higher order effects must be considered
to completely fit the displacement, values of $\beta$ can
nonetheless be evaluated in the range of $10^{13}\,{\rm m}^{-2}$.
Also note that such a more complete derivation \cite{Landau}
would lead to the emergence of odd harmonics of motion with an
amplitude at least $\varepsilon \lesssim 10^{-3}$ times smaller
than the fundamental oscillation at $\Omega_{\rm p}$.

Stable solutions of eqs. \eqref{Eq:S_x_bista} and
\eqref{Eq:epsilon} correspond to the upper and lower branches of
a bistable hysteresis cycle. As the network analyzer only sweeps
frequencies upward, curves in Fig. \ref{faisceau} (left) explore
the upper branch until they suddenly fall down to the lower
branch. To sweep frequency either up or down, we make use of a
signal generator together with a spectrum analyzer in max-hold
configuration (see setup {\it b} in Fig.~\ref{Montage}). Both
sweeps are shown in Fig.~\ref{faisceau} (right) and clearly
exhibits an hysteresis cycle.

\begin{figure}[b]
\includegraphics[width=8.6cm]{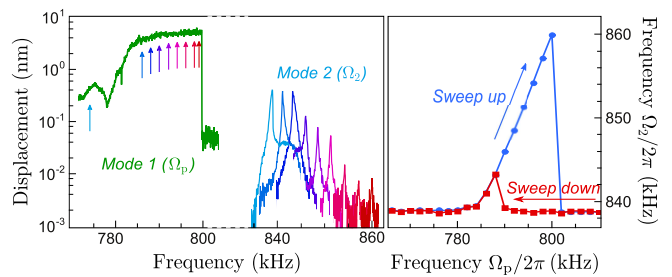}
\caption{\label{mode_shifting} Intermodal coupling: actuation of
mode 1 in its nonlinear regime (left curve) induces a frequency
shift of the resonance of mode 2 (central curves) when the
actuation frequency $\Omega_{\rm p}$ of mode 1 is increased along
the upper branch (arrows). Right curves display the bistable
behavior of the resonance frequency $\Omega_{\rm s}$ of mode 2
when $\Omega_{\rm p}$ is either swept up or down. The
nonlinearity strength of mode 1 is comparable to the one experienced in curve {\it e} in
Fig.~\ref{Eq:S_x_bista}.}
\end{figure}

The nonlinearity arises from additional stress induced by large
displacements and we expect nonlocal effects as intrinsic elastic
parameters such as Young's modulus of the membrane may be affected
\cite{PhysRevB.84.134305, venstra:151904}. We have thus investigated
the influence of one mode (mode 1) on the natural frequency of
another one (mode 2), using both the signal generator and the
network analyzer as shown in Fig. \ref{Montage}.

An upward frequency sweep is first produced by the signal generator
with a 20-dBm power to actuate the first mode in its nonlinear
regime. In order to explore different points on the upper branch of
the nonlinearity, the sweep is stopped at different frequencies
$\Omega_{\rm p}$, represented by the arrows in Fig.
\ref{mode_shifting} (left). While the drive of mode 1 remains at
that frequency, we monitor the linear response of mode 2 using the
network analyzer with a 0-dBm actuation power. Central curves in
Fig. \ref{mode_shifting}, obtained for increasing actuation
frequencies $\Omega_{\rm p}$, clearly show that the resonance
frequency $\Omega_{\rm s}$ of mode 2 is shifted as the displacement
amplitude  of mode 1 gets larger and larger. Considering that mode 2
has a quality factor of $6\,000$, the mode is actually displaced 170
times its width. We also explore the lower branch of the nonlinear
response of mode 1 using an initial downward frequency sweep (Fig.
\ref{mode_shifting} right), thus retrieving the bistable behavior of
mode 1 on the resonance-frequency shift of mode 2.

\begin{figure}
\includegraphics[width=8.6cm]{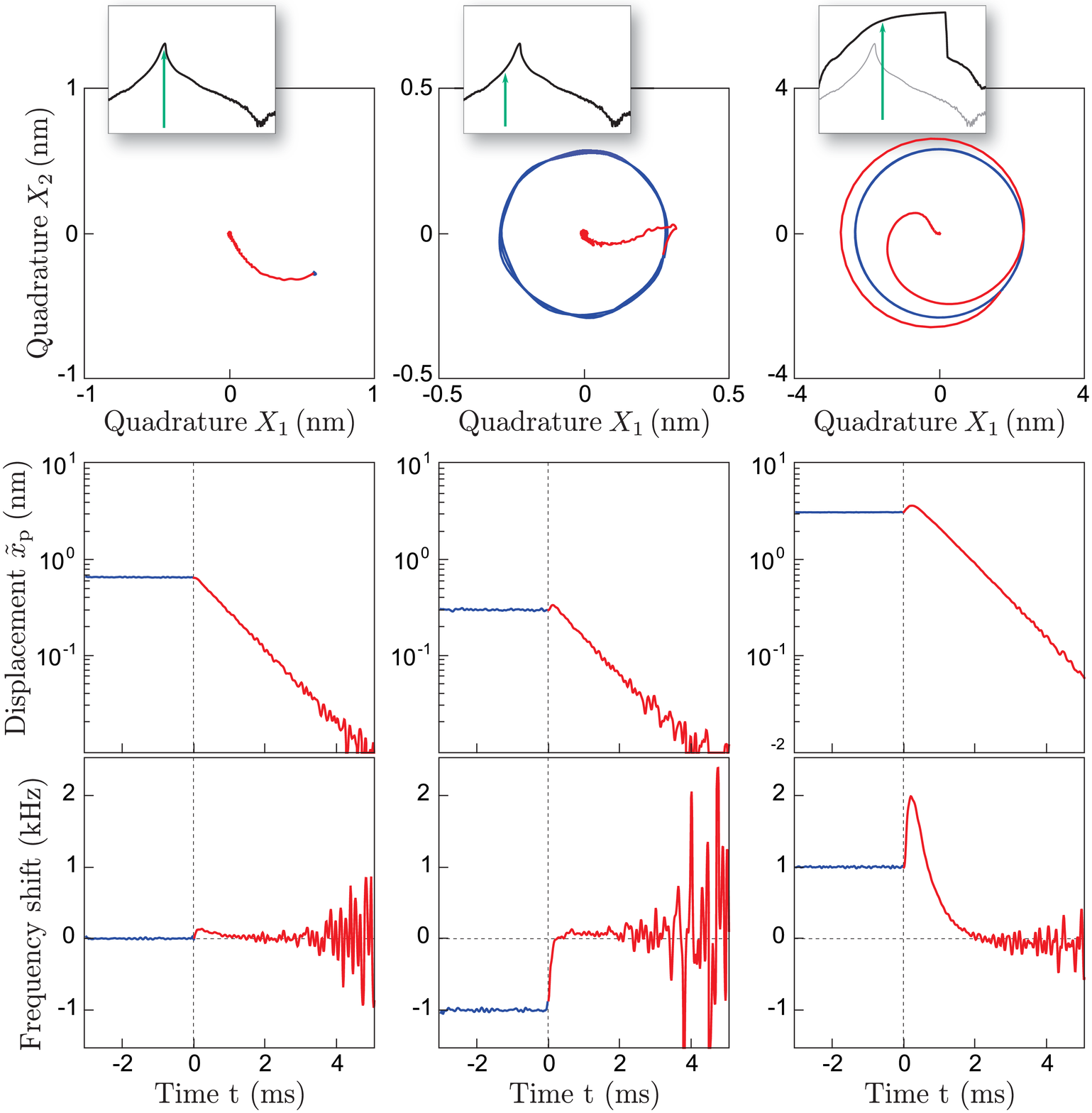}
\caption{\label{IQ} Time evolution in phase-space for the forced
(blue) and free (red) regimes. Three cases are investigated, as
sketched on top of each column, from left to right: linear regime
actuated at resonance (5~dBm at $1\,057\ {\rm kHz}$), out of
resonance (9~dBm at $1\,056\ {\rm kHz}$), and nonlinear regime
(29~dBm at $1\,058\ {\rm kHz}$). Curves from top to bottom represent
the phase-space trajectories and the amplitude and frequency
evolutions.}
\end{figure}

As the effective elastic parameters of the membrane are altered by
the nonlinearity, one may also expect a modification of its
dynamics, which we have investigated using a ring-down technique. We
have thus monitored the motion of the membrane in phase space,
defining the slowly varying quadratures $X_1$ and $X_2$ as:
\begin{equation}
x_\mathrm{p}(t) = X_1(t) \cos(\Omega_0 t) + X_2(t) \sin(\Omega_0 t).
\end{equation}
$X_1$ and $X_2$ are readily accessible using the spectrum analyzer
in IQ demodulation mode. Figure~\ref{IQ} shows the forced regime and
the ring-down decay of the mechanical mode for different actuation
levels: in the linear regime at its natural frequency $\Omega_0$
(left column), in the linear regime out of resonance (central
column), and in the nonlinear regime (right column). The membrane is
actuated for times $t<0$, with an upward frequency sweep to get the
mode on the upper branch in the nonlinear case, then the actuation
is stopped by a fast switch at time $t=0$.

The phase-space trajectories (top curves) first show the forced
regime which appears as blue circles or as a single stationary point
in the resonant case (left), and second the ring-down regime
displayed as red spirals which reduce to an almost straight line
down to the origin when the motion is close to resonance.

>From these data we can infer the time evolutions of both the
amplitude and frequency of motion (middle and bottom curves). In
contrast to other systems such as graphene \cite{Eichler}, the
amplitude decay time is the same in the different actuation
regimes: $\tau=1/\Gamma\simeq 0.56\ {\rm ms}$, corresponding to a
mechanical quality factor $Q\simeq 3\,700$. This tends to confirm
that the damping in these resonators does not depend on the
effective Young modulus and is indeed mostly due to surface
effects \cite{Antoni:11}.

Yet the frequency behavior drastically changes between the linear
and nonlinear regimes. In the former case, even for a non-resonant
actuation, the frequency instantaneously jumps back to its natural
value when the actuation is released, as can also be inferred from
the straight line trajectories in phase space. This result is
actually expected for a linear harmonic oscillator, as its free
evolution for $t>0$ only depends on the initial position and speed
at $t=0$, and not on the previous forced actuation frequency. In the
nonlinear case, the frequency hops towards the top of the upper
bistability branch and then slowly relaxes down to the natural
frequency as the nonlinearity fades out. This can be understood from
eq. \eqref{Eq:S_x_bista} which shows that for a given oscillation
amplitude $\tilde{x}_\mathrm{p}$, the preferred oscillation
frequency is $\Omega_0\left(1+\varepsilon/2\right)$ \cite{Landau}. As the oscillation amplitude and then the nonlinear
coefficient $\varepsilon$ [eq. \eqref{Eq:epsilon}] exponentially
decrease with time, one thus expects the same behavior for the free
oscillation frequency.

\begin{figure}[b]
\includegraphics[width=8.6cm]{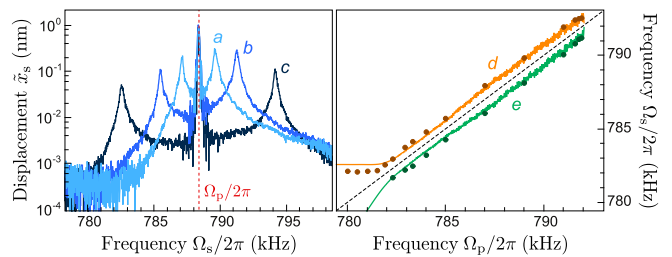}
\caption{\label{mixing} Mechanical response of a mode pumped in
the nonlinear regime (frequency $\Omega_{\rm p}$) and probed by a
weak actuation (frequency $\Omega_{\rm s}$). Spectra obtained
with pump actuation powers of 10, 20, and 30~dBm (curves {\it a}
to {\it c}) exhibit two resonances symmetrically disposed around
the pump frequency $\Omega_{\rm p}$, as demonstrated by curves
{\it d} and {\it e}: dots are experimental values of the two
resonance frequencies obtained for a pump actuation of 5~dBm
(leading to a nonlinearity strength similar to curve {\it c}
in Fig.~\ref{Eq:S_x_bista}), solid curves are theoretical fits.}
\end{figure}

To confirm this interpretation, we have studied the mechanical
response of the mode to a weak probe excitation while it is
simultaneously set in the nonlinear regime by a strong pump
actuation. The setup is actually similar to the one used for the
two-mode actuation (Fig. \ref{mode_shifting}), except that both pump
and probe are actuating the same mechanical mode, close to its
natural frequency $\Omega_0/2\pi\simeq 782\ {\rm kHz}$ for the
membrane used here. Figure ~\ref{mixing} (left) shows the spectra
$\tilde{x}_\mathrm{s}[\Omega_{\rm s}]$ obtained when the probe
frequency $\Omega_{\rm s}$ is scanned around the pump frequency
$\Omega_{\rm p}$, for different pump actuation powers. We clearly
observe two symmetrical lorentzian peaks whose frequency difference
increases with the pump power.

Using a spectrum analyzer instead of a network analyzer, we
actually checked that the two resonances {\it simultaneously}
appear: when actuated at a frequency $\Omega_{\rm s}$, the
pump-probe combination leads to the concomitant generation of a
mechanical response at frequencies $\Omega_{\rm s}$ and
$2\Omega_{\rm p}-\Omega_{\rm s}$, symmetrically disposed on both
sides of the pump frequency.

This behavior can be understood from the Duffing equation
\eqref{Eq:Duffing}, adding a second source term
$\alpha_\mathrm{s}(t) = 2 \tilde{\alpha}_\mathrm{s} \cos
(\Omega_\mathrm{s} t)$ related to the probe. The cubic term in
Duffing equation is then responsible for the mixing of pump and
probe signals, in a way similar to a $\chi^{(3)}$ nonlinearity in
optics. Indeed, writing the total displacement in presence of the
probe as the sum of the unperturbed displacement $x_\mathrm{p} (t)$
due to the pump only [eq. \eqref{Eq:S_x_bista}] and an additional
displacement $x_\mathrm{s} (t)$, we obtain the following equation
(assuming
$\left|\tilde{x}_\mathrm{s}\right|\ll\left|\tilde{x}_\mathrm{p}\right|$):
\begin{equation}
\ddot{x}_\mathrm{s}(t) + \Gamma \dot{x}_\mathrm{s}(t) +
\Omega_0^2 \left[ 1 + 3 \beta x_\mathrm{p}^2(t) \right]
x_\mathrm{s}(t) = \alpha_{\rm s} (t). \label{Eq:x_s}
\end{equation}
As $x_{\rm p}(t)$ is oscillating at frequency $\Omega_{\rm p}$,
the nonlinear term $x_{\rm p}^2(t)$ contains both a static term
and a fast oscillation at $2\Omega_{\rm p}$. The former is
responsible for a modification of the linear mechanical response
at frequency $\Omega_{\rm s}$ whereas the latter is formally
similar to phase conjugation in optics and generates a response
at $2\Omega_{\rm p}-\Omega_{\rm s}$. From \eqref{Eq:x_s} one gets
the following coupled equations for the two Fourier components of
$x_{\rm s}$ at frequencies $\Omega_{\rm s}$ and
$2\Omega_{\rm p}-\Omega_{\rm s}$:
\begin{eqnarray}
\zeta[\Omega_\mathrm{s}]\ \tilde{x}_\mathrm{s}
[\Omega_\mathrm{s}] + \varepsilon\,\Omega_0^2\
\tilde{x}_\mathrm{s}^\star [2 \Omega_\mathrm{p} -
\Omega_\mathrm{s}] &=& \tilde{\alpha}_\mathrm{s},
\label{eq_sonde1}\\
\zeta[2 \Omega_\mathrm{p} - \Omega_\mathrm{s}]\
\tilde{x}_\mathrm{s}^\star[2 \Omega_\mathrm{p} -
\Omega_\mathrm{s}] + \varepsilon\,\Omega_0^2\
\tilde{x}_\mathrm{s} [\Omega_\mathrm{s}] &=& 0, \label{Eq:sonde2}
\end{eqnarray}
where $\zeta[\Omega] = \Omega_0^2 \left(1+ 2 \varepsilon \right)-
\Omega^2 - i \Gamma \Omega_0$, and we have assumed for
simplicity $\tilde{x}_{\rm p}$ real. We thus obtain two
phase-conjugate lorentzian responses with resonance frequencies:
\begin{equation}
\Omega_{{\rm s}+} = \Omega_0 \left(1 + \varepsilon \right),\quad
\Omega_{{\rm s}-}=2\Omega_\mathrm{p}-\Omega_{{\rm s}+}.
\label{Eq:conjugue_freq}
\end{equation}
Note the difference between $\Omega_{{\rm s}+}$ and the resonance frequency
$\Omega_0\left(1+\varepsilon/2\right)$ in response to a single
actuation [eq. \eqref{Eq:S_x_bista}], actually equivalent to the difference obtained between self- and cross-phase modulations in Kerr media\cite{SPMXPM}.

Dots in Fig. \ref{mixing} (right) show frequencies $\Omega_{{\rm s}+}$  and $\Omega_{{\rm s}-}$ deduced from spectra similar to the ones displayed on Fig.
\ref{mixing} (left), as a function of the pump frequency
$\Omega_{\rm p}$. Curves {\it d} and {\it e} are
theoretical predictions obtained using eqs.
\eqref{Eq:conjugue_freq} with $\varepsilon$ estimated from the
experimental nonlinear behavior [eq. \eqref{Eq:S_x_bista} and
Fig. \ref{faisceau}]. At low pump frequency $\Omega_{\rm
p}\lesssim \Omega_0$, the membrane is almost in a linear regime
and we only observe one peak close to the natural frequency
$\Omega_0$. At higher frequency, the two resonances are
symmetrically located around $\Omega_{\rm p}$ (dashed line), in
good agreement with theoretical predictions.

We have demonstrated nonlinear effects in the dynamics
of PhC nanomembranes: bistability, intermodal tuning and
the generation of phase conjugate modes. Pump-probe experiments
have in particular allowed us to underline the non-local character of the
Young modulus modification. Such nonlinear behaviours are intrinsic to
nano- and microscale systems\cite{rhoads:034001}, and have dramatic consequences
not only in the fields of optomechanics and
nonlinear dynamics, but in quantum optics as well, as intermodal
coupling can be used to perform QND measurements
of a signal mode by monitoring a meter one \cite{Grangier,PhysRevLett.105.117205,PhysRevB.70.144301}. Optomechanical
systems can also be
used for classical or quantum information processing, either by taking
advantage of their long mechanical coherence times to store
information in a mechanical excitation \cite{Weis}, or by using their
nonlinear character for non volatile memories \cite{nnano.2011.180}.

{\bf Acknowledgements}

The authors gratefully thank Alexios Beveratos and Izo Abram for
fruitful scientific discussions, Luc Le Gratiet and Gr\'egoire
Beaudoin for assistance in the fabrication of the nanomembranes.
This research has been partially funded by the FP7 Specific
Targeted Research
Projects QNems, and by the C'Nano \^{I}le-de-France project Naomi. \\

\end{document}